# cgmquantify: Python and R packages for comprehensive analysis of interstitial glucose and glycemic variability from continuous glucose monitor data

Brinnae Bent[1], Maria Henriquez[2], Jessilyn Dunn[1,3]

[1]Department of Biomedical Engineering, Duke University, Durham, North Carolina

[2]Department of Statistical Science, Duke University, Durham, North Carolina

[3]Department of Biostatistics and Bioinformatics, Duke University Medical Center, Durham, North Carolina

TAGS

Open source, Python, R, bioinformatics, data analysis, glucose, glucose variability, continuous glucose monitoring, type I diabetes, type II diabetes, diabetes, prediabetes, Open APS

SUMMARY

Continuous glucose monitoring (CGM) systems provide real-time, dynamic glucose information by tracking interstitial glucose values throughout the day (typically values are recorded every 5 minutes). CGMs are commonly used in diabetes management by clinicians and patients and in research to understand how factors of longitudinal glucose and glucose variability relate to disease onset and severity and the efficacy of interventions. CGM data presents unique bioinformatic challenges because the data is longitudinal, temporal, and there are nearly infinite possible ways to summarize and use this data. There are over 20 metrics of glucose variability, no standardization of metrics, and little validation across studies. Here we present open source python and R packages called cgmquantify, which contains over 20 functions with over 25 clinically validated metrics of glucose and glucose variability and functions for visualizing longitudinal CGM data. This is expected to be useful for researchers and may provide additional insights to patients and clinicians about glucose patterns.

INTRODUCTION

Continuous glucose monitoring (CGM) systems provide real-time, dynamic glucose information by tracking interstitial glucose values throughout the day. CGMs are commonly used in diabetes management, with 1.2 million diabetic patients using a CGM[18]. CGM use has be associated with improved glycemic control in adults with type 1 diabetes[14]. These devices have been used extensively by the T1D community, including in the Open Artificial Pancreas System Project (OpenAPS)[10], a project developed to create a patient-implemented closed loop system between a CGM and an insulin pump.

CGM data is commonly provided from CGM manufacturers as either raw glucose values (in a .csv format) or in summary reports that utilize proprietary methods to plot and summarize glucose statistics (e.g. Dexcom Clarity currently shows overall mean glucose, standard deviation of glucose, time in range, and hypoglycemia risk and daily minimum, maximum, mean, and standard deviation of glucose). Because these algorithms are proprietary, they cannot be properly validated by clinical researchers[4]. Additionally, the provided glucose summaries are extremely limited and do not usually contain any information about an important clinical metric, glycemic variability.

Glycemic variability, also known as glucose variability, is an established risk factor for hypoglycemia[15] and has been shown to be a risk factor in diabetes complications[12]. Glucose variability can be found in over 26,000 publications indexed in PubMed at the time of this publication and is a significant metric in clinical research[8]. Over 20 metrics of glucose variability have been identified (Table 1), which makes it difficult to examine and compare results across numerous research studies analyzing and drawing conclusions about glucose variability.

There is a need for an open source resource with algorithms that are utilized and validated in clinical research studies. This would enable standardized glucose variability metrics and the ability to compare findings from studies that utilize different metrics of glucose variability. This resource should be available in an open source programming language with a low barrier to entry to encourage researchers, clinicians, and patients alike to explore CGM data.

Previous open-source resources have been implemented in Excel[6] and R[16,19]. There is currently no comprehensive resource for CGM data in Python, the third most common programming language used globally and the leading language among newcomers[9]. Additionally, previous implementations of open source CGM data analysis have limited metrics of glucose variability. Further, these methods are typically developed for a specific purpose and are therefore not extensible (e.g. do not have simple functions so users can customize their metrics and visualizations).

We have developed a package written and published in Python under the MIT license and a package written and published in R under the MIT license. The packages, both named cgmquantify, contain over 20 functions with more than 25 metrics summarizing glucose and glucose variability. There is also includes code for visualizing CGM data in both packages. Both packages are available in the Digital Biomarker Discovery Pipeline (DBDP)[1], the open source software platform for digital biomarker discovery. The python package is available under the Python Package Index (PyPI) (https://pypi.org/project/cgmquantify/). Source code can be found at https://github.com/DigitalBiomarkerDiscoveryPipeline/cgmquantify. The R package is available in the Comprehensive R Archive Network (CRAN) (https://CRAN.R-project.org/package=cgmquantify) and source code can be found at https://github.com/DigitalBiomarkerDiscoveryPipeline/cgmquantify-R.

METHODS

cgmquantify is a Python package and an R package composed of 20+ functions with over 25 clinically validated metrics of glucose and glucose variability, as shown in Table 1. Customizable visualizations (Figure 1, Figure 2) are also included as easy to implement functions. cgmquantify is version controlled through GitHub and PyPI or CRAN. This allows for single-line installation in either language. Source code and an extensive user guide are available on GitHub to facilitate ease of use and enable customization based on user needs. Issue tracking on GitHub is monitored closely by the Digital Biomarker Discovery Pipeline to allow for rapid feedback. Tests are available in GitHub under the tests subdirectory to allow for manual testing of all functions.

RESULTS

We have included import functions to format data for use with the cgmquantify package. These functions currently support Dexcom CGM devices, with plans to add additional import functions for other CGM manufacturers, including Medtronic and Abbott. Our user guide also outlines how one can easily format data to make any data input compatible with the functions in cgmquantify. Functions are available for all the commonly studied glucose and glucose variability metrics (Table 1). Additionally, functions for data visualization of the longitudinal CGM data are provided. These visualizations are easily customizable. We have also implemented a function that enables LOWESS smoothing over the CGM data (Figure 1).

DISCUSSION

cgmquantify is a package that simplifies the process of calculating metrics and thus allows for easy comparison across different research studies that use different metrics summarizing glucose and glucose variability. Functions have been developed using equations from clinically validated research studies so users can compare their results to previous findings. The cgmquantify package is easily implemented with a one-line installation and an extensive user guide in both the python and R languages. Detailed documentation facilitates modification of existing code for customization of input and visualizations. This package also has the ability to build a community of developers to contribute to the literature in this burgeoning field.

This is a much-needed resource for the community of researchers, clinicians, and patients using CGM. Currently, little is understood about the relationships between glucose and glucose variability metrics from CGM data and relationships to diseases including but not limited to prediabetes, T2D, and severity of symptoms in T1D. As more researchers and clinicians start looking to CGM data to answer these questions, the need for a standardized resource in a nearly ubiquitous programming language is necessary. As we have seen with the Open APS community, analysis of CGM data is not limited to researchers and clinicians but includes patients themselves[5]. By providing this as an open source resource, we hope to encourage patients to interact with their own data, determine personalized insights, and make meaningful contributions to the digital health landscape.

FUTURE IMPLEMENTATIONS

Future contributions will include additional import functions customized to all the CGM manufacturers, including but not limited to Medtronic and Abbott. We are exploring methods to incorporate food logs into visualizations of CGM data.

CODE AVAILABILITY

The cgmquantify python package is available under the Python Package Index (PyPI) (https://pypi.org/project/cgmquantify/). Source code can be found at https://github.com/DigitalBiomarkerDiscoveryPipeline/cgmquantify. The cgmquantify R package is available in the Comprehensive R Archive Network (CRAN) (https://CRAN.R-project.org/package=cgmquantify) and source code can be found at https://github.com/DigitalBiomarkerDiscoveryPipeline/cgmquantify-R. We encourage others to expand on our ideas and contribute their own glucose and glucose variability metrics to cgmquantify. We have documentation for contributing available in our User Guide.

REFERENCES

1.  Bent B, Wang K, Grzesiak E, et al. The Digital Biomarker Discovery Pipeline: An open source software platform for the development of digital biomarkers using mHealth and wearables data. *J Clin Transl Sci*. 2020;(11):1-28. doi:10.1017/cts.2020.511.

2.  Cho P, Bent B, Wittmann A, et al. Expanding the Definition of Intraday Glucose Variability. In: *Diabetes*. American Diabetes Association; 2020.

3.  eAG/A1C Conversion Calculator | American Diabetes Association.

https://professional.diabetes.org/diapro/glucose_calc. Accessed February 15, 2020.

4. Goldsack J, Coravos A, Bakker J, Bent B, Dowling AV, Fitzer-Attas C, Godfrey A, Godino JG, Gujar N, Izmailova E, Manta C, Peterson B, Vandendressche BV, Wood WA, Wang KW DJ. Verification, Analytical Validation, and Clinical Validation (V3): The Foundation of Determining Fit-for-Purpose for Biometric Monitoring Technologies (BioMeTs). *JMIR Prepr #17264*. 2020. https://preprints.jmir.org/preprint/17264.

5. De Groot M, Drangsholt M, Martin-Sanchez F, Wolf G. Single Subject (N-of-1) Research Design, Data Processing, and Personal Science. *Methods Inf Med*. 2017;56(06):416-418. doi:10.3414/ME17-03-0001.

6. Hill NR, Oliver NS, Choudhary P, Levy JC, Hindmarsh P, Matthews DR. Normal reference range for mean tissue glucose and glycemic variability derived from continuous glucose monitoring for subjects without diabetes in different ethnic groups. *Diabetes Technol Ther*. 2011;13(9):921-928. doi:10.1089/dia.2010.0247.

7. Kovatchev B. Glycemic Variability: Risk Factors, Assessment, and Control. *J Diabetes Sci Technol*. 2019;13(4):627-635. doi:10.1177/1932296819826111.

8. Kovatchev BP. Metrics for glycaemic control — from HbA1c to continuous glucose monitoring. *Nat Rev Endocrinol*. 2017;13(7):425-436. doi:10.1038/nrendo.2017.3.

9. Krill P. Python popularity reaches an all-time high | InfoWorld. https://www.infoworld.com/article/3401536/python-popularity-reaches-an-all-time-high.html. Published 2019. Accessed February 12, 2020.

10. OpenAPS.org – #WeAreNotWaiting to reduce the burden of Type 1 diabetes. https://openaps.org/. Accessed February 15, 2020.

11. Rawlings RA, Shi H, Yuan LH, Brehm W, Pop-Busui R, Nelson PW. Translating glucose variability metrics into the clinic via continuous glucose monitoring: A graphical user interface for diabetes evaluation (CGM-GUIDE©). *Diabetes Technol Ther*. 2011;13(12):1241-1248. doi:10.1089/dia.2011.0099.

12. Service FJ. Glucose variability. *Diabetes*. 2013;62(5):1398-1404. doi:10.2337/db12-1396.

13. Suh S, Kim JH. Glycemic variability: How do we measure it and why is it important? *Diabetes Metab J*. 2015;39(4):273-282. doi:10.4093/dmj.2015.39.4.273.

14. Tamborlane W V., Beck RW, Bode BW, et al. Continuous Glucose Monitoring and Intensive Treatment of Type 1 Diabetes. *N Engl J Med*. 2008;359(14):1464-1476. doi:10.1056/NEJMoa0805017.

15. Umpierrez GE, P. Kovatchev B. Glycemic Variability: How to Measure and Its Clinical Implication for Type 2 Diabetes. *Am J Med Sci*. 2018;356(6):518-527. doi:10.1016/j.amjms.2018.09.010.

16. Vigers T, Chan CL, Snell-Bergeon J, et al. cgmanalysis: An R package for descriptive analysis of continuous glucose monitor data. Bethin KE, ed. *PLoS One*.

2019;14(10):e0216851. doi:10.1371/journal.pone.0216851.

17. Wojcicki JM. 'J'-index. A new proposition of the assessment of current glucose control in diabetic patients. *Horm Metab Res*. 1995;27(1):41-42. doi:10.1055/s-2007-979906.

18. Wolinsky H. Continuous Glucose Monitors: The Long and Winding Road To Acceptance, Coverage | Managed Care magazine. Managed Care. https://www.managedcaremag.com/archives/2018/7/continuous-glucose-monitors-long-and-winding-road-acceptance-coverage. Published 2018. Accessed February 15, 2020.

19. Zhang XD, Zhang Z, Wang D. CGManalyzer: An R package for analyzing continuous glucose monitoring studies. *Bioinformatics*. 2018;34(9):1609-1611. doi:10.1093/bioinformatics/btx826.

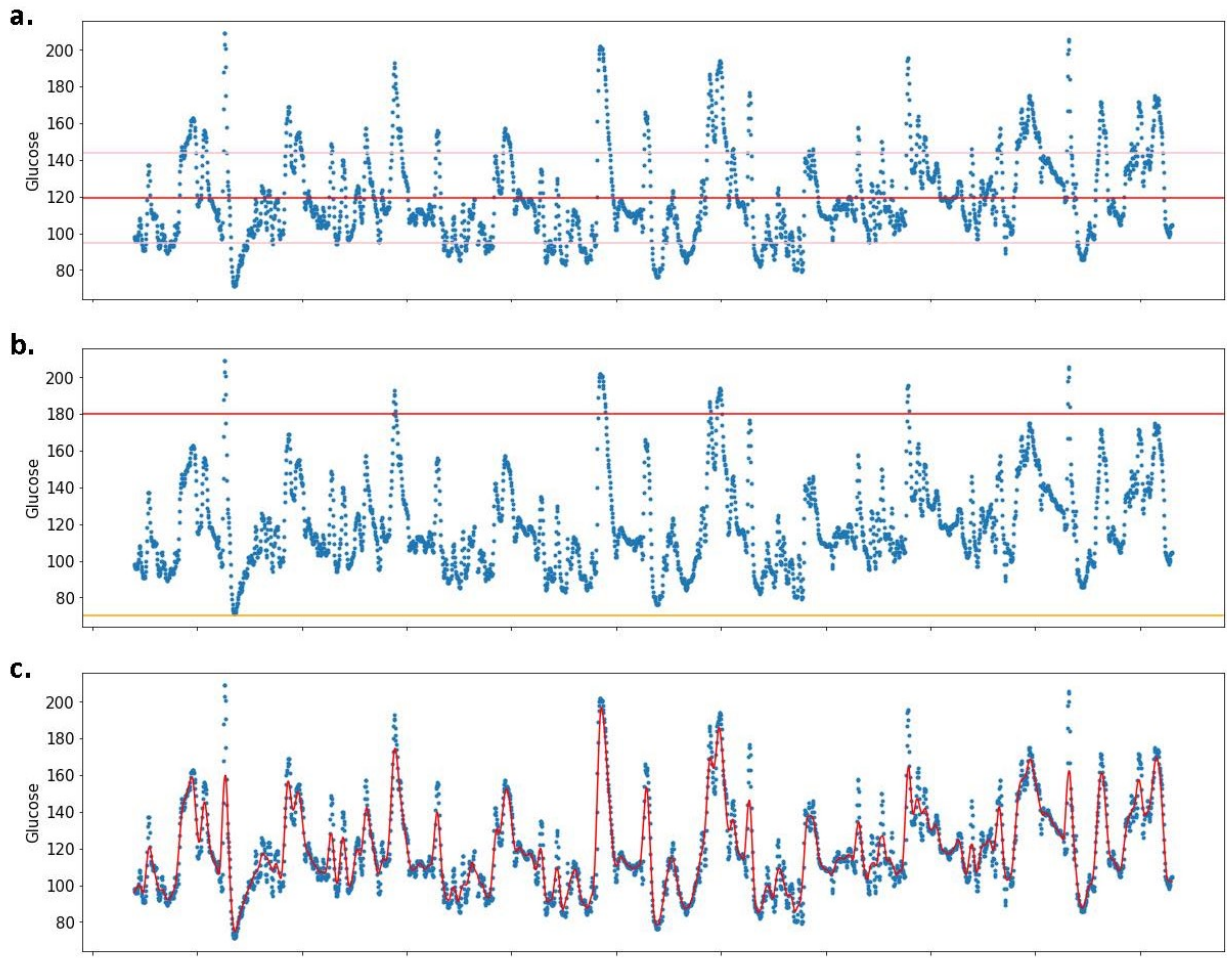

**Figure 1. Visualizing longitudinal CGM data with the cgmquantify Python package.** Shown are a visualization with indicators of 1 SD from the mean and the mean glucose level (a), a visualization with indicators of hyperglycemic (>180 mg/dL glucose) and hypoglycemic (<70 mg/dL glucose) (b), and a plot with LOWESS smoothing of the glucose data (c).

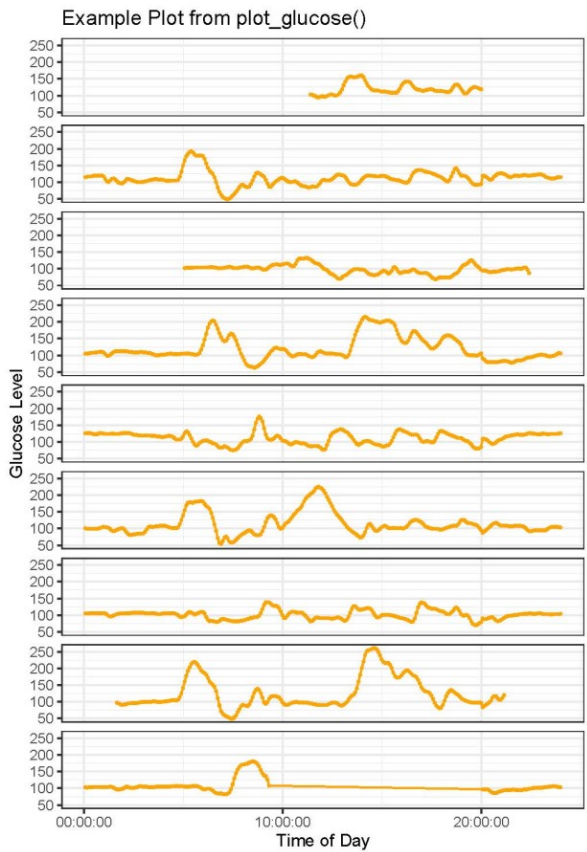

**Figure 2. Visualizing longitudinal CGM data with the cgmquantify R package.** Shown is a visualization available in the cgmquantify R package that enables visualization of CGM data by time of day for each day specified.

**Table 1.** Glucose and Glucose Variability Metrics

| Metric | Description | Equation |
|---|---|---|
| interdaySD[8,12] | Interday standard deviation of glucose | $\sigma_{interday} = \sqrt{\frac{\sum(G_i - \mu)^2}{N}}$ <br> Where N = total days, G = glucose value |
| interdayCV[8] | Interday coefficient of variation of glucose | $CV_{interday} = \frac{\sigma_{interday}}{\mu}$ |
| intradaySD Mean[8] | Intraday standard deviation of glucose (mean across all days) | $\sigma_{intraday_{mean}} = \sum_N \sigma_i$ |
| intradayCV Mean[8] | Intraday coefficient of variation of glucose (mean across all days) | $CV_{intraday_{mean}} = \sum_N CV_i$ |
| intradaySD Median[2] | Intraday standard deviation of glucose (median across all days) | $\sigma_{intraday_{median}} = median(\sigma_i)$ |
| intradayCV Median[2] | Intraday coefficient of variation of glucose (median across all days) | $CV_{intraday_{median}} = median(CV_i)$ |
| intradaySD Standard Deviation[2] | Intraday standard deviation of glucose (standard deviation across all days) | $\sigma_{intraday_{standard\ deviation}} = SD(\sigma_i)$ |
| intradayCV Standard Deviation[2] | Intraday coefficient of variation of glucose (standard deviation across all days) | $CV_{intraday_{standard\ deviation}} = SD(CV_i)$ |
| CONGA24[11,12] | Continuous overall net glycemic action over 24 hours | $CONGA24 = SD(|G_t - G_{t-24hours}|)$ |
| GMI[19] | Glucose management indicator | $GMI = 3.31 + (0.02392 * \mu\ (mg/dL))$ |
| HBGI[8,12] | High Blood Glucose Index | $\sum \frac{r_l}{n}$ <br> $r_l = 22.7 * (f(G_i))^2$ if $f(G_i) \leq 0, r_l = 0$ otherwise <br> $f(G_i) = \ln(G_i)^{1.084} + 5.381$ |
| LBGI[8,12] | Low Blood Glucose Index | $\sum \frac{r_h}{n}$ <br> $r_h = 22.7 * (f(G_i))^2$ if $f(G_i) > 0, r_l = 0$ otherwise <br> $f(G_i) = \ln(G_i)^{1.084} + 5.381$ |
| ADRR[8] | Average Daily Risk Range, assessment of total daily glucose variations within risk space | $ADRR = \frac{\sum_{all\ days}(LR^j + HR^j)}{N\ days}$ <br> where $LR^j = \max(r_l)$ and $HR^j = \max(r_h)$ |
| J-index[12,17] | Measure of both the mean level and variability of glycemia | $J = 0.001 * (\mu + \sigma)^2$ |
| MAGE[12,13] | Mean amplitude of glucose excursions (default = 1SD) | 1. Local minima/maxima determined <br> 2. Assess max/min pairs against SD |

| | | | |
|---|---|---|---|
| | | | 3. If difference from min to max > SD, mean measure is retained<br>4. Otherwise excluded<br>5. Troughs are retained and summed |
| MGE[12] | Mean of glucose outside range (default = 1SD) | | $\mu_{glucose\ outside\ \#\ SD\ of\ mean}$<br><br>Where # is set, default is 1 SD |
| MGN* | Mean of glucose inside range (default = 1SD) | | $\mu_{glucose\ inside\ \#\ SD\ of\ mean}$ |
| MODD[8,11] | Mean of daily differences in glucose | | $MODD = \dfrac{\sum |G_t - G_{t-24hours}|}{total\ matched\ observations}$ |
| TIR[7] | Time spent in range (minutes), default = 1SD | | $TIR = \sum_N time\ inside\ \#\ SD\ of\ mean$ |
| TOR[7] | Time spent outside range (minutes), default = 1SD | | $TOR = \sum_N time\ outside\ \#\ SD\ of\ mean$ |
| POR[16] | Percent of time spent outside range | | $POR = \dfrac{TOR}{total\ time} \times 100\%$ |
| PIR[16] | Percent of time spent inside range, default = 1SD | | $PIR = \dfrac{TIR}{total\ time} \times 100\%$ |
| eA1c[3] | Estimated A1c (according to American Diabetes Association) | | $eA1c = \dfrac{(46.7 + \mu)}{28.7}$ |
| meanG[11] | Mean glucose over all days | | $\mu = \dfrac{\sum_N \bar{x_i}}{N}$ |
| medianG[16,19] | Median glucose over all days | | median $(Glucose)$ |
| minG[16] | Minimum glucose over all days | | min $(Glucose)$ |
| maxG[16] | Maximum glucose over all days | | max $(Glucose)$ |
| Q1G[16] | First quartile glucose value over all days | | first quartile $(Glucose)$ |
| Q3G[16] | Third quartile glucose value over all days | | third quartile $(Glucose)$ |

* indicates previously unidentified metrics of glucose variability that are similar to clinically validated metrics